\documentclass[twocolumn,showpacs,preprintnumbers,pra,floatfix,superscriptaddress,amsmath,amssymb]{revtex4}
\usepackage{graphicx}
\usepackage[latin1]{inputenc}

\newcommand{\depr}{\left(\mathbf{r}\right)}

\newcommand{\deprt}{\left(\mathbf{r},t\right)}

\newcommand{\be}{ \begin{equation} }
\newcommand{\ee}{ \end{equation} }
\newcommand{\bea}{\begin{eqnarray} }
\newcommand{\eea}{\end{eqnarray} }
\newcommand{\bean}{\begin{eqnarray*} }
\newcommand{\eean}{\end{eqnarray*} }
\newcommand{\br}{\mathbf{r}}
\newcommand{\bx}{\mathbf{x}}

\begin{document}

\title{Generation of nonground-state Bose-Einstein condensates by modulating
atomic interactions}

\author{E. R. F. Ramos}\email{edmir@ursa.ifsc.usp.br}
\author{E. A. L. Henn}
\author{J. A. Seman}
\author{M. A. Caracanhas}
\author{K. M. F. Magalh\~{a}es}
\affiliation{Instituto de F\'{i}sica de S\~{a}o Carlos, Universidade
de S\~{a}o Paulo, Caixa Postal 369, 13560-970, S\~{a}o Carlos-SP
Brazil}
\author{K. Helmerson}
\affiliation{Joint Quantum Institute, NIST and University of
Maryland, College Park, Maryland 20742, USA}
\author{V. I. Yukalov}
\affiliation{Bogolubov Laboratory of Theoretical Physics, \\ Joint
Institute for Nuclear Research Dubna 141980, Russia \\ [5mm]}
\author{V. S. Bagnato}

\affiliation{Instituto de F\'{i}sica de S\~{a}o Carlos, Universidade
de S\~{a}o Paulo, Caixa Postal 369, 13560-970, S\~{a}o Carlos-SP
Brazil}

\begin{abstract}
A technique is proposed for creating nonground-state Bose-Einstein
condensates in a trapping potential by means of the temporal
modulation of atomic interactions. Applying a time-dependent
spatially homogeneous magnetic field modifies the atomic
scattering length. A modulation of the scattering length excites
the condensate, which, under special conditions, can be
transferred to an excited nonlinear coherent mode. It is shown
that a phase-transition-like behavior occurs in the time-averaged
population imbalance between the ground and excited states. The
application of the technique is analyzed and it is shown that the
considered effect can be realized for experimentally available
condensates.
\end{abstract}


\maketitle

\section{Introduction}

Bose-Einstein condensation (BEC) is an effect of a very broad
interest, touching on a variety of physical topics and creating,
sometimes unexpected, links among different physical subjects. Many
exciting possibilities have been investigated in recent years and
are discussed in books~\cite{pitaevskii} and review
articles~\cite{{review},{andersen},{girardeau}}. The Feshbach
resonance technique, which enables a variation of the scattering
length via a magnetic field, is one of the most promising tools for
manipulating the properties of quantum degenerate gases. This
technique, e.g., has made it possible to tune an unstable system
into a stable one~\cite{{saito},{koch}}, to form molecular
condensates~\cite{molecular}, and to investigate the nonlinear
dynamics of a BEC with a time-dependent scattering
length~\cite{{dynamics},{adhikari}}.

In the present letter, we show that the temporal modulation of the
scattering length can also be used for generating nonground-state
condensates of trapped atoms. Such states are described by nonlinear
topological coherent modes and can be excited by a resonant
modulation of the trapping potential~\cite{{nonground},{yyb}}. We
advance an alternative way for exciting the coherent modes of a
trapped BEC by including an oscillatory component in the scattering
length. The main idea is to superimpose onto the BEC a uniform
magnetic field with a small amplitude time variation. Due to the
Feshbach resonance effect, such an oscillatory field creates an
external perturbation in the system, coherently transferring atoms
from the ground to a chosen excited coherent state. The feasibility
of the experimental implementation of this phenomenon for available
atomic systems is demonstrated.

The coherent states of a trapped BEC are described by the solutions
to the Gross-Pitaevskii equation (GPE). To transfer the BEC from the
ground to a nonground state, it is necessary to apply a
time-dependent perturbation, at a frequency close to the considered
transition. As a result~\cite{{nonground},{yyb}} the resonantly
excited condensate becomes an effective two-level system. The
external fields considered in the previous
works~\cite{{review},{nonground},{yyb},{eta}} were formed by
spatially inhomogeneous alternating trapping potentials. Now, we
consider a very different situation represented by a spatially
homogeneous time-oscillating magnetic field, which can be easily
implemented with present experimental techniques.

\section{Modulation of scattering length}

The GPE, describing a zero-temperature weakly interacting Bose gas, is
given by
 \be
  i\hslash\frac{\partial\Phi}{\partial t} =
\left[-\frac{\hslash^2}{2m_0}\nabla^2 + U_{trap}\depr + A_s
  \left|\Phi\right|^2\right]\Phi,
  \label{eq:gpe}
  \ee
where $U_{trap}\depr$ is the trapping potential, the interaction
strength $A_s=4\pi(N-1)\hslash^2 a_s/m_0$, $a_s$ the s-wave
scattering length, $m_0$ the atomic mass and N is the number of
condensed atoms. In the presence of a spatially uniform magnetic
field, $a_s$ near a Feshbach resonance is given by the well known
relation
 \be
 a_s = a_{nr} \left(1-\frac{\Delta}{B-B_{res}}\right),
 \label{eq:as}
 \ee
where $a_{nr}$ is a non-resonant scattering length, $B_{res}$ is the
value of the magnetic field where the resonance in $a_s$ occurs, and
$\Delta$ is the corresponding resonance width. Let us consider the
time-dependent magnetic field
 \be
 B(t) = B_0 + b \cos(\omega t),
 \ee
with $|b| \ll |B_0-B_{res}|$. In such a case, Eq.(\ref{eq:as}) can
be expanded to first order as
 \be
 a_s(t) \simeq a_0 + a \cos(\omega t),
 \label{eq:asap}
 \ee
where
 \be
 a_0 = a_{nr} \left(1-\frac{\Delta}{B_0-B_{res}}\right)
 \ \ \ \
 a =  \frac{a_{nr} \, b \,\Delta }{\left(B_0-B_{res}\right)^2} \, .
 \label{eq:a0 e a}
 \ee
The scattering length then possesses an oscillatory component around the
average value.

Combining Eq.(\ref{eq:asap}) and Eq.(\ref{eq:gpe}), one gets the GPE
with the additional oscillatory term $V=V\deprt$. With the notation
$H~=~H_0 + V$, one has
 \bea
  &H\Phi=i\hslash \; \partial\Phi / \partial t,&
  \label{eq:gpet}\\
 \nonumber\\
 &H_0= \displaystyle{-\frac{\hslash^2}{2m_0}\nabla^2 + U_{trap}\depr + A_0
 \left|\Phi\right|^2,\label{eq:h0}}&\\
 \nonumber\\
 &V=A \cos (\omega t) \left|\Phi\right|^2\label{eq:v},&
 \eea
$$
A_0=\frac{4\pi (N-1) \hslash^2}{m_0}\,a_{0} \, ,
$$
$$
A=\frac{4\pi (N-1) \hslash^2}{m_0}\,a \, .
$$

Solving Eq.(\ref{eq:gpet}), we keep in mind that the frequency
$\omega$ is chosen to be  close to the transition frequency between
the ground and an excited mode. We starting considering as the total
wavefunction a linear combination of a complete set of modes as
follows
 \be
\Phi\deprt=\sum_j c_j(t) \phi_j(\br) e^{-iE_jt/\hslash},
 \label{eq:phij}
 \ee
where $\phi_j(\br)$ are stationary solutions for the equation
$H_0\phi_j=E_j\phi_j$, with eigenenergies $E_j$. Then, was proved
before~\cite{{review},{nonground}} the only relevant terms that
survive are two modes connected by the modulating perturbation
(\ref{eq:v}). In this case, the total wavefunction (\ref{eq:phij})
can be represented, in a good approximation, by
 \be
 \Phi\deprt=c_0(t) \phi_0(\br) e^{-iE_0t/\hslash}+c_p(t)
\phi_p(\br) e^{-iE_pt/\hslash},
 \label{eq:2modes}
 \ee
where the label $0$ refers to ground state and $p$ to an excited
state. We investigate the time evolution of the BEC, with the
initial condition, where all atoms are in the ground state, i.e.,
$c_0(0)=1$ and $c_p(0)=0$. Our aim is to study the transfer between
the fractional mode populations of the ground and the excited
states. Using Eq.(\ref{eq:2modes}) in Eq.(\ref{eq:gpet}), one
obtains a set of differential equations for the coefficients
$c_0(t)$ and $c_p(t)$,
\begin{subequations}
 \bea
i\hslash\frac{dc_0}{dt} & = &
A_0 \left|c_p\right|^2 c_0 \left( 2 I_{0,p,0} - I_{0,0,0} \right)
\nonumber \\
& + & \frac{A}{2}e^{i\Delta\omega t} \left( \left|c_p\right|^2 c_p
I_{0,p,p} + 2\left|c_0\right|^2 c_p
I_{0,0,p}\right) \nonumber\\
 & + & \frac{A}{2} e^{-i\Delta\omega t} c_p^* c_0^2 I_{p,0,0} \; ,
 \label{eq:c0}
 \eea
 \bea
  i\hslash\frac{dc_p}{dt} & = &
A_0 \left|c_0\right|^2 c_p \left( 2 I_{p,0,p} - I_{p,p,p} \right)
\nonumber\\
&+& \frac{A}{2}e^{-i\Delta\omega t} \left( \left|c_0\right|^2 c_0
I_{p,0,0} +2\left|c_p\right|^2 c_0
I_{p,p,0}\right) \nonumber\\
&+ & \frac{A}{2}e^{i\Delta\omega t} c_0^* c_p^2 I_{0,p,p} \; ,
\label{eq:cp}
 \eea
 \label{eq:dc0dcp}
\end{subequations}
where the integral $I_{j,k,l}$ is defined as
 \be
 I_{j,k,l} = \int \phi_j^* |\phi_k|^2 \phi_l d\br.
 \ee
In deriving the latter equations, two assumptions, whose
mathematical basis has been described in detail in
Refs.~\cite{{review},{nonground},{yyb}}, are made. First, the time
variation of $c_0(t)$ and $c_p(t)$ are to be much slower than the
the exponential oscillations with the transition frequency
$\omega_{p0} = (E_p-E_0)/\hslash$. This condition is fulfilled, when
the amplitudes $A_0 I_{j,k,l}$ and $A I_{j,k,l}$ are smaller than
$\hslash \omega_{p0}$. The second is the resonance condition, when
the external alternating field connects only the two chosen
nonlinear states. Another point concerns damping due to collisions
between particles in the desired modes or collisions with the
thermal cloud. Although the oscillation time for populations takes
tens of trap periods, this time is much smaller than the lifetime of
a typical BEC or a vortex state~\cite{{vortex1},{vortex1}}. So, we
expect that damping occurs but not as a dominate process. Thus, we
have left out the damping effect for this model.

Another important aspect is that the total number of atoms do not
vary on time, but the number in each state does. This variation is
taking into account in Eqs. (\ref{eq:dc0dcp}) since these equations
depend on the population of each state, represented by $|c_0|^2$ and
$|c_p|^2$. However, the modes $\phi_j$ in equation (\ref{eq:phij})
are stationary solutions of Equation (\ref{eq:h0}) when all atoms
are in state $j$. Thus, if there is a variation in the atom number
of some state, there is a variation in the wave function that
represents this state. So, the total wave function should be written
in the form
 \be
 \Phi\deprt=\sum_j d_j(t) \phi'_j(\br,t),
 \label{eq:phijprime}
 \ee
where the number dependence is inserted in the time dependence. In
this way, the population of a state $j$ would be given by
$|d_j(t)|^2$ and not by $|c_j(t)|^2$, since the expansions
(\ref{eq:phij}) and (\ref{eq:phijprime}) are different. However, in
the case of our study, the system is in a weak-coupling regime,
i.e., $g$ is small, so the variation of the wave function can be
neglected and the population of a state $j$ can be given by
$$ n_j(t) \approx |c_j(t)|^2.$$

\section{Application to a cylindrically symmetric trap}

We consider a cylindrically symmetric harmonic trap
 \be
 U_{trap} = \frac{m_0}{2}(\omega_r^2 r^2 + \omega_z^2 z^2),
 \ee
and use the optimized perturbation theory, as discussed in Ref.
\cite{{review},{nonground},{yyb}}, for finding the modes $\phi_0$ and
$\phi_p$. It is convenient to define the dimensionless variables
\begin{eqnarray}
 \psi(\bx)=l_r^{3/2}\phi\depr, \ \ &
\ \ \displaystyle{H(\psi)= \frac{H(\phi)}{\hslash\omega_r} },  \nonumber\\ \nonumber\\
\displaystyle{x_r = \frac{r}{l_r}, } \ \ & \ \ \displaystyle{x_z =
\frac{z}{l_r} }, \nonumber
\end{eqnarray}
 \be
g_0 = 4\pi (N-1)\frac{ a_0}{l_r} \; , \quad \lambda =
\frac{\omega_z}{\omega_r} \; , \quad \delta = \frac{(\omega -
\omega_{p0})}{\omega_r} \; ,
 \label{eq:g0}
 \ee
where $l_r=\sqrt{\hslash/m_0\omega_r}$. Using the fourth-order
Runge-Kutta method~\cite{maple}, we calculate the time evolution of
the coefficients $c_0(t)$ and $c_p(t)$ for different values of the
detuning $\delta$ and external-field amplitude. The functions
$c_0(t)$ and $c_p(t)$ define the fractional mode populations.

\begin{figure}[!h]
\centering
\includegraphics[scale=1.5]{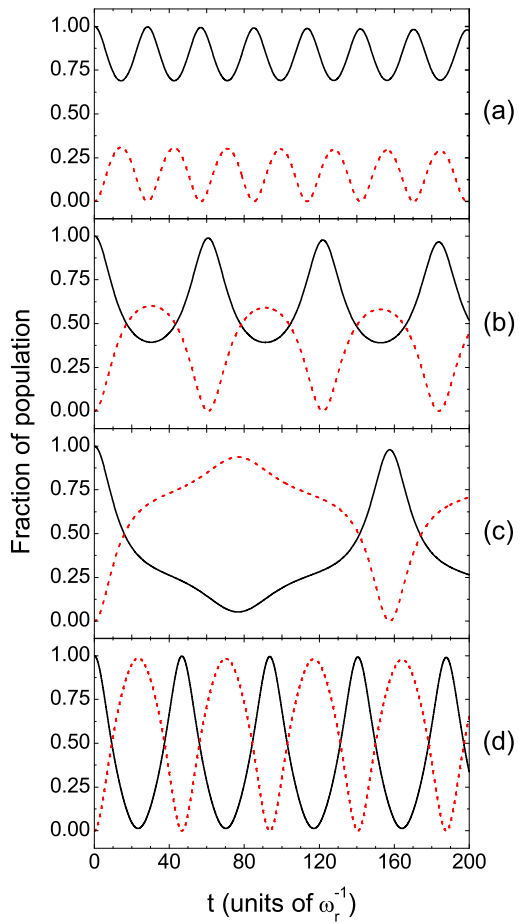}\\
\caption{(Color online) Populations of the ground state $n_0$ (solid
line) and excited state $\{100\}$ $n_p$ (dashed line) as a function
of time for $\lambda=0.2$ and $g_0=70$ with (a) $a/a_0=0.7$ and
$\delta=0$; (b) $a/a_0=0.7$ and $\delta=0.04$; (c) $a/a_0=0.72$ and
$\delta=0.04$; (d) $a/a_0=1$ and $\delta=0.04$} \label{fig:pop}
\end{figure}

For an excited mode, we take the radial dipole state $\{100\}$,
which is the lowest energy state above the ground state $\{000\}$.
Here $\{nmk\}$ implies the notation for the three quantum numbers,
where $n$, $m$ and $k$ refer to radial, azimuthal and axial mode
numbers respectively. Fig.\ref{fig:pop}, where $\lambda=0.2$ and
$g_0=70$, shows the time evolution of the mode populations $n_0$ and
$n_p$ for different values of the detuning $\delta$ and $a/a_0$,
which is given by Eq. (\ref{eq:a0 e a}). The chosen parameters
correspond to typical experimental setups and are easily controlled
in a laboratory.
The solutions demonstrate different behaviors of the state
populations. For $a/a_0 = 0.7$ and $\delta=0$, Fig.\ref{fig:pop}(a),
the populations display small oscillation amplitudes, with a
considerably larger population in the ground state. Increasing the
detuning to $\delta=0.04$ results in the behavior shown in
Fig.\ref{fig:pop}(b). Although on average atoms stay longer in the
ground state, for some intervals of time $n_p$ is larger than $n_0$.
Changing the amplitude to $a/a_0=0.72$ and maintaining
$\delta=0.04$, as in Fig.\ref{fig:pop}(c), yields a very different
temporal behavior. Atoms now stay longer in the excited state rather
than in the ground state. The shape of the functions shows the
inherent nonlinearity of the system. If the amplitude is increased
further to $a/a_0=1$, with $\delta=0.04$, as in Fig.
\ref{fig:pop}(d), the system shows a full population inversion. For
certain times, when the mode population fully migrates from the
ground $\{000\}$ to the excited $\{100\}$ state, it is possible to
have a pure condensate in the coherent topological excited mode.

A convenient way to quantify the population behavior is through the
introduction of an order parameter $\eta$, defined as the difference
between the time-averaged populations  for both
states~\cite{{eta},{ramos}},
 \be
 \eta=\bar{n}_0-\bar{n}_p.
 \label{eq:eta}
 \ee
Here, the average of each population is performed over the full cycle
of an oscillation.

\begin{figure}[!h]
\centering
\includegraphics[width=0.53\textwidth=1]{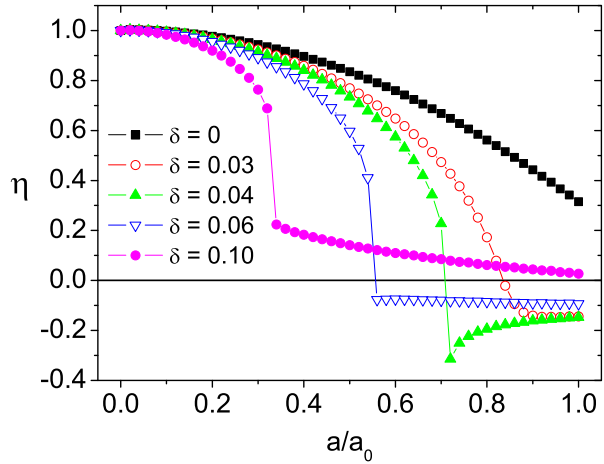}\\
\caption{(Color online) Order parameter $\eta$ as a function of the
ratio $a/a_0$, for different detunings showing a phase-transition
like behavior.} \label{fig:eta}
\end{figure}

The above order parameter $\eta$ displays a nontrivial behavior when
the ratio $a/a_0$ is modified. For different detunings, the
variation of $\eta$ as a function of the ratio $a/a_0$ is presented
in Fig.\ref{fig:eta}.
The variation of $\eta$ vs. $a/a_0$ can be smooth, when $\delta \leq
0.03$, or can show sudden changes, when $\delta>0.03$. Smaller
values of $a/a_0$ keep atoms preferably in the ground state. For
some critical value of $a/a_0$, $\eta$ becomes negative, which means
that BEC stays longer with a larger population in the excited state
than in the ground state. This situation is ideal for detecting the
formation of topological modes. Step like behavior observed on
$\eta$ is not surprising for a nonlinear system. For example, the
classical driven anharmonic oscillator \cite{landau} exhibits a
shift on its resonance curve, which we also observe in our system
\cite{ramos pla}. This shift for a large driven strength can result
in a bistable behavior. In our system that appears as a step like
response with $\eta$. This bistability appears only for positive
detunings, as observed in our calculation. Negative detunings
follows a smooth curve without step occurrence.

An important question is the feasibility of the experimental
creation of such coherent modes. The main parameter here is $a/a_0$.
It shows us that this value is strongly dependent on the Feshbach
resonance width $\Delta$, characteristic of each type of atoms. For
systems with small $\Delta$, the required value of $a/a_0$ for
obtaining the transition in $\eta$ will occur only for $B_0 \approx
B_{res}$. In this case, the necessary condition $b\ll B_{res}-B_0$
can only be fulfilled for very small values of $b$, which creates an
extra difficulty with the present techniques of magnetic field
control~\cite{{strecker02},{errico07}}. As an experimentally
realistic example, let us consider the case of $b =
0.1(B_{res}-B_0)$ and $a/a_0=0.8$, which corresponds to the atomic
parameters listed in Table \ref{tab}. Setting $g_0=70$,
$\omega_r=2\pi\times 120\,$Hz, we obtain a value of $b$ of $0.9\,G$
for $^{85}$Rb, and of $0.02\,G$ for $^{87}$Rb, which would be
difficult to control. On the other hand, $10\,G$, for $^7$Li, and
$4.55\,G$, for $^{39}$K, are values that can be realized with
present technical capabilities~\cite{{strecker02},{errico07}}.

Let us consider now a condensate containing $10^5$ $^7$Li atoms in a
trap with radial frequency $\omega_r=2\pi\times 120\,$Hz, $g_0=70$,
and $\lambda=0.2$. With these conditions, the transition frequency
is $\omega_{p0}=\omega_{100,0}=2\pi\times 209.6\,$Hz. Also, together
with the information from Table \ref{tab}, Eqs. (\ref{eq:a0 e a})
and (\ref{eq:g0}), we obtain the bias magnetic field
$B_0~=~632.5\,G$ and the amplitude $b = 11.68\,a/a_0$. Then, we
observe the critical values for $b$ ranging from $4\,G$ for a
detuning $\omega-\omega_{p0}=2\pi\times 21\,$Hz to $10.5\,G$, for
$\omega-\omega_{p0}=2\pi\times 6.3\,$Hz. Such oscillating amplitudes
correspond to less than $1\%$ of the total bias field $B_0$.

\begin{table}
  \centering
\begin{tabular}{c | c c c c c c c}
\hline\hline Atom& $B_{res}(G)\;$ & $\; \Delta(G)$ & $\; a_{nr}\;$ &
$\; B_0(G)\;$ &
$\; b(G)\;$ & $\; a_0$ \;& $\; N (\times 10^4)$ \\
\hline
$^{85}$Rb & 155.0 & 10.7 & -443 & 164.4 & -0.9 & 63 & 0.2\\
$^{87}$Rb & 1007.34 & 0.17 & 100 & 1007.53 & 0.02 & 11 & 0.9\\
$^{7}$Li & 735 & -113 & -27.5 & 636 & 10 & 3.9 & 9.3\\
$^{39}$K & 403.4 & -52 & -23 & 357.9 & 4.55 & 3.3 & 4.7\\
\hline\hline
\end{tabular}
\caption{Amplitudes $B_0$ and $b$ of the magnetic field for four
different species of atoms . We set $g_0 = 70$, $\lambda=0.2$,
$a/a_0~=~0.8$ and $b = 0.1(B_{res}-B_0)$. Scattering length is
expressed in units of the Bohr radius. Data for $^{85}$Rb were
obtained in Ref.~\cite{claussen96}; for $^{87}$Rb,
Ref.~\cite{marte02}; for $^{7}$Li, Ref.~\cite{strecker02}; for
$^{39}$K, Ref.~\cite{errico07}.}\label{tab}
\end{table}


A final point to be addressed concerns losses introduced by
collisions, specially near a Feshbach resonance. With an
off-resonant magnetic field, the dominant loss mechanism is a three
body collision, whereas close to the resonance, the molecular
formation dominates the atom loss mechanism~\cite{stenger99}.
Although the fields considered in Table \ref{tab} and in the $^7$Li
example above are within the resonance linewidth, they are far
enough to the resonance, then we consider three body collision as
the main loss mechanism. In the case of $^7$Li, that the resonance
linewidth is large, as we are in the border of this linewidth, the
loss rate is very small, as shown in Ref.~\cite{strecker02}. So, we
can neglect the atom loss, specially because we apply the magnetic
field in a short period of time.

\section{Discussion}

In conclusion, we have shown that using a magnetic modulation field,
applied to a trapped Bose-condensed gas, it is possible to transfer
the atomic population from the ground state to an excited state,
producing a nonground-state condensate. The time averaged population
imbalance between the ground and excited states represents an order
parameter, which demonstrates an interesting behavior as a function
of the modulation amplitude. Depending on the detuning, the behavior
of $\eta$ can be either smooth or rather abrupt. This is the
consequence of the strong nonlinearity of the interactions. For some
range of detunings, $\eta$ becomes negative above a critical value
of the modulation amplitude. This occurs because of the population
inversion realized during the process of the mode excitation. Larger
detunings and out-of-resonance modulations keep the population in
the ground state and no population inversion is observed. Numerical
calculations, accomplished for $^7$Li atoms, show that the values
for the amplitude and modulation of the bias field are within
realistic experimental conditions.


The justification of the approach for generating nonlinear
coherent modes by the resonant modulation of an alternating field
has been explained in detail in our previous publications
\cite{{nonground},{yyb},{eta},{yukalov03a},{yukalov03b}}. It would
be unreasonable to repeat here all this justification in full.
However, for the convenience of interested readers, we recall in
brief some important points.

When a system is subject to the action of a time-dependent
external field $V(t)$, alternating with a frequency $\omega$, then
there exist two principally different situations, depending on the
temporal alternation being either slow or fast \cite{zaslavsky}.
Respectively, there occur two different physical cases.

If the temporal variation is slow, such that the alternation
frequency $\omega$ is much smaller than  the characteristic system
energy $E$, then one says that the perturbation is adiabatic
\cite{zaslavsky}. For a quantum-mechanical system, the energy $E$
is an eigenvalue of the system Hamiltonian $H$. When the latter
varies slowly in time, so that the variation is adiabatic, then
the adiabatic picture is in order, with the eigenvalues $E(t)$ and
related eigenfunctions $\varphi(t)$ slowly varying in time. For
such an adiabatic variation, the notion of nonlinear coherent
modes is not of much use.

A principally different situation occurs, when the temporal
variation is fast, so that $\omega$ is in resonance with one of
the transition frequencies $\omega_{ij}\equiv(E_i-E_j)/\hbar$. It
is exactly this case that is considered in the present paper. Then
the solutions of the stationary equation
$H_0\varphi_j=E_j\varphi_j$ define the nonlinear coherent modes
$\varphi_j$. Constructing a closed linear envelope, spanning the
total set $\{\varphi_j\}$, one gets the Hilbert space ${\cal
H}\equiv{\rm Span}\{\varphi_j\}$, with the typical property
\cite{richtmyer} that any function from ${\cal H}$ can be expanded
over the set $\{\varphi_j\}$. The latter set forms a total basis
\cite{{yyb},{zhidkov}}, over which the solution of Eq. (6) can be
expanded as in Eq. (9). Looking for the solution of Eq. (6) in the
form of expansion (9), we employ the method of the parameter
variation. Separating in expansion (9) the time-dependent
coefficient function into two factors, the fastly oscillating
exponential $\exp(-iE_jt/\hbar)$, and the slowly varying envelope
$c_j(t)$, we can use the averaging techniques \cite{bogolubov} and
the scale separation approach \cite{{yukalov96},yukalov00}. A
similar technique in optics is called the slowly-varying amplitude
approximation [32]. Substituting expansion (9) into Eq. (6) and
involving the averaging techniques [29--31], we obtain Eqs. (11)
for $c_n(t)$. This procedure is the same as has been done in Refs.
[10,11].

In this way, one should not confuse the adiabatically slow
variation of an external field, when the mode profiles and
energies are certainly changing in time, and the fast resonant
field oscillation, when the fractional mode populations vary
between the stationary coherent modes. In the latter case, the
solution to Eq. (6) can be represented in form (9) and the
averaging techniques are applicable, yielding Eqs. (11) for the
coefficient functions. The situation here is equivalent to the
resonant excitation of an atom, as is discussed in Refs. [24,25].
In the same way as for an atom, the external resonant field should
not be too strong in order not to disturb the energy-level
classification. For this, it is sufficient to take the modulation
amplitude $b$ in Eq. (3) appropriately small, which is always
possible.

It is important to stress that the averaging technique, employed
for deriving the equations for the fractional mode coefficients
$c_n(t)$, is a well justified method based on rigorous
mathematical theorems [29]. Therefore, this method does not
require some other justifications.

Moreover, the results, obtained by means of the averaging
techniques, when treating the generation of the nonlinear coherent
modes by a resonant alternating field, have been thoroughly
compared with the results of the direct numerical simulation of
the Gross-Pitaevskii equation in Refs. [33,34]. Both ways of
treating the problem have been found to be in a very good
quantitative agreement.

In a real experiment, of course, a modulation, even being
perfectly resonant, might, nevertheless, lead to heating, when the
neighboring nonresonant levels become essentially involved in the
process. This means that such a resonant mode generation can be
effective only during a finite time. This problem has been
analyzed in detail earlier [11,35], where it has been found that,
taking the modulation amplitude sufficiently small, it is feasible
to avoid heating during the lifetime of atoms inside a trap. For
instance, the heating time was estimated [11,35] to be of order
$10-100$ s.

The nonground-state nonlinear coherent modes are often termed
topological, since the wavefunctions of different modes possess
drastically different spatial shapes, with a different number and
location of zeros. As an example, we can recall the wavefunctions
of the first excited modes for atoms in a harmonic trap, found in
Refs. [11,24]. These functions $\psi_{nmk}(x_r,\varphi,x_z)$,
represented in dimensionless units of Eq. (14), are labelled by
the quantum numbers $n$, $m$, and $k$ and are the functions of the
dimensionless cylindrical variables $x_r,\varphi$, and $x_z$. The
ground-state wavefunction is
$$
\psi_{000} = \left ( \frac{u_{000}^2v_{000}}{\pi^3} \right )^{1/4}
\; \mbox{e}^{-\left ( u_{000} x_r^2 + v_{000} x_z^2 \right )/2} \;
.
$$
The radial dipole mode has the form
$$
\psi_{100} = \left ( \frac{u_{100}^2v_{100}}{\pi^3} \right )^{1/4}
\left ( u_{100} x_r^2 - 1 \right ) \; \mbox{e}^{-\left ( u_{000}
x_r^2 + v_{000} x_z^2 \right )/2} \; .
$$
The vortex mode is also a possible stationary solution
$$
\psi_{010} = u_{010} \left ( \frac{v_{010}}{\pi^3} \right )^{1/4}
x_r \; \mbox{e}^{i\varphi} \; \mbox{e}^{-\left ( u_{000} x_r^2 +
v_{000} x_z^2 \right )/2} \; .
$$
And the axial dipole mode is
$$
\psi_{001} = \left ( \frac{4u_{001}^2v_{001}^3}{\pi^3} \right
)^{1/4} x_z \; \mbox{e}^{-\left ( u_{000} x_r^2 + v_{000} x_z^2
\right )/2} \; .
$$
The quantities $u_{nmk}$ and $v_{nmk}$ here are defined by the
optimization conditions and depend on all system parameters (see
calculational details in Refs. [11,24]). As is evident, the
spatial shapes of atomic clouds, described by $|\psi_{nmk}|^2$,
are principally different for different quantum numbers.

If the scattering length is modulated according to Eq. (4), then
the resulting Eq. (6) acquires the additional term (8) playing the
role of a modulating field. Therefore the mechanism of generating
nonground-state modes is the same for both the cases, whether the
additional term is caused by the trap modulation or by the
scattering-length modulation. In any case, the general structure
of Eqs. (6) to (8) is similar for both these setups.

The main difference between the methods of generating nonlinear
coherent modes by the trap modulation or by the scattering-length
modulation is as follows. In the method of the trap modulation, by
choosing the appropriate spatial dependence of the modulating
field $V({\bf r},t)$, it is possible to generate any type of
modes. While, in the method of the scattering-length modulation,
only those modes can be excited, for which the integrals in Eq.
(12) are nonzero. For example, in the case of cylindrically
symmetric harmonic trap with potential (13), the integrals
$I_{0,p,p}$, $I_{0,0,p}$, $I_{p,0,0}$, and $I_{p,p,0}$, in which
$p$ implies a set $\{ n,m,k\}$, vanish for the vortex mode
$\psi_p=\psi_{010}$ and for the axial dipole mode
$\psi_p=\psi_{001}$. Hence, in these cases, the vortex and the
axial dipole modes cannot be excited by modulating the scattering
length. However, the radial dipole mode $\psi_{100}$ can be
generated by this method, since the related integrals (12) are
nonzero. This is why, we have considered here exactly this case. A
way out of the problem, which would allow for the generation of
other modes, including the vortex mode, could be by employing a
nonsymmetric trapping potential.

\begin{acknowledgments}
Authors thank R. G. Hulet and G. Roati for providing us with
experimental data on Feshbach resonances, and L. Tomio for fruitful
suggestions. This work was supported by CAPES, CNPq and FAPESP.
\end{acknowledgments}

\appendix

\end{document}